\documentclass[runningheads]{llncs}
\usepackage[T1]{fontenc}
\usepackage{graphicx,verbatim}
\usepackage{amsmath, amssymb}
\usepackage{dsfont}
\usepackage{booktabs}

\begin{document}
\title{From Explainable to Explained AI: Ideas for Falsifying and Quantifying Explanations
}
%

\author{Yoni Schirris\inst{1} \and
Eric Marcus\inst{1} \and
Jonas Teuwen\inst{1} \and 
Hugo Horlings\inst{1} \and 
Efstratios Gavves\inst{2}}

\authorrunning{Y. Schirris et al.}
%
\institute{Netherlands Cancer Institute \and University of Amsterdam
}



\maketitle              
\begin{abstract}

Explaining deep learning models is essential for clinical integration of medical image analysis systems. A good explanation highlights if a model depends on spurious features that undermines generalization and harms a subset of patients or, conversely, may present novel biological insights. Although techniques like GradCAM can identify influential features, they are measurement tools that do not themselves form an explanation. 
We propose a human–machine-VLM interaction system tailored to explaining classifiers in computational pathology, including multi-instance learning for whole-slide images. Our proof of concept comprises (1) an AI-integrated slide viewer to run sliding-window experiments to test claims of an explanation, and (2) quantification of an explanation's predictiveness using general-purpose vision-language models. The results demonstrate that this allows us to qualitatively test claims of explanations and can quantifiably distinguish competing explanations. This offers a practical path from explainable AI to explained AI in digital pathology and beyond. Code and prompts are available at \url{https://github.com/nki-ai/x2x}.


\end{abstract}

\begin{figure}[htb]
\includegraphics[width=\textwidth]{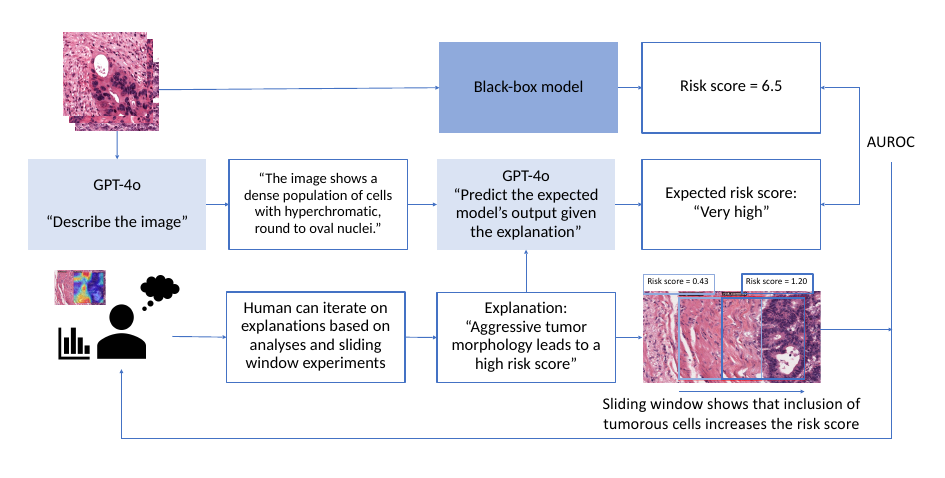}
\caption{Overview of our proposed evaluation framework for explanations. Sliding window experiments test the change in the black-box model output logit as specific features are included in the field of view, to test if the explanation could predict this. To quantitatively test if the hypothesized black-box explanation can accurately predict the output of the black-box model, the VLM follows an explanation to score a patch, which is compared to the black-box model output logit.} 
\label{fig:overview}
\end{figure}

\section{Introduction}

The application of deep learning (DL) in medical image analysis has become ubiquitous across tasks and domains \cite{balkenende2022application,mcgenity2024artificialpathologysystematicreview} and is increasingly being implemented in the clinic. However, the black-box nature of DL often leaves these models unexplained. Most DL models are correlational learners, often leading to shortcut learning \cite{geirhos2020shortcut}. The model then uses spurious features to make predictions that may lead to unwanted outcomes if implemented in the clinic. Examples are available for the predictions of the response to chemotherapy from breast magnetic resonance images \cite{saha2018machinejorencommentedon,brunekreef2024letter} where the model remembered patient anatomy; in survival prediction from multimodal data including computational pathology \cite{chen2022pancancermultimodalsurvival,howard2023multimodal} where medical center patterns themselves were predictive of survival; for COVID-19 prediction from CT images \cite{degrave2021ai} where the model focused on technical artifacts on a scan; and medical center memorization by pathology foundation models \cite{jongmarcusteuwen2025fomomedicalcenter}. These examples show that high-performance metrics on a validation dataset are not proof of a good model that has learned the underlying biology and causal structure of the task. We believe that providing an explanation of what the model does, rigorously testing, and iterating upon the explanation can help prevent such outcomes and guide researchers to overcome these issues during model development, and is essential for the clinical applicability of these models.

There is an increasing research interest in explainable AI (XAI) to explain how black-box models work in medical imaging \cite{van2022explainablereview,borys2023explainablesaliencyreview}. For example, saliency methods like GradCAM \cite{selvaraju2017gradcam} highlight the image's most influential parts for the current prediction. These visualizations are often called ``the explanation'' \cite{borys2023explainablesaliencyreview}. However, this visualization is only a measurement, which itself does not form an explanation. For example, the microscopic image of a striped jellybean inside a cell (mitochondria) told the viewer in the 19\textsuperscript{th} century nothing of its role in energy production or its intricate evolutionary origins. In medical image analysis, another standard practice is to perform post-hoc correlative analyses of the model's output with clinical variables on the validation set \cite{wulczyn2021interpretable}. Such studies investigate (in)dependence with (un)wanted variables and, combined with other XAI measurements, are used to generate an initial explanation of what the model might do. However, such an explanation is rarely tested further, as would be standard practice in the sciences. One reason for this omission is the complex collaboration required between highly trained medical and AI experts (e.g., \cite{reitsam2024converging} tests the biological validity of the explanation of \cite{wulczyn2021interpretable}). We believe that another important reason is that \textit{the generation and validation of good explanations of deep learning models is an open problem, lacking conceptual frameworks, practical guidelines, and toolkits}. 

\paragraph{\textbf{Problem}}: The absence of ideas and methods for the falsification and quantification of explanations for AI models limits the individual researcher to confidently provide good explanations to discover and prevent unwanted actions by the model.

\vspace{0.1cm}

To this end, we present a prototype for human-in-the-loop testing of components of a black box model's explanation in section \ref{sec:viewer} and automated quantitative evaluation of such an explanation using a general-purpose vision-language model (VLM) in section \ref{sec:evaluation}. Although this work focuses on a risk prediction model for colorectal cancer in computational pathology as described in section \ref{sec:definition-explanation}, these ideas are transferrable to other tasks and domains. 

Before we can investigate tools to challenge and quantify the predictive power of explanations, however, we should agree on what an ``explanation'' is.

\section{What is an Explanation?} \label{sec:definition-explanation}

We define an \textit{explanation} of a DL model as a hypothesized natural language description of what the model does, such that it is predictive of the model's output for samples that fall within the distribution that the explanation is hypothesized for. 

Imagine a cow-camel classifier \cite{arjovsky2019invariant} where the training set consists of cows in meadows and camels in deserts. Naively, we think the model distinguishes a cow from a camel independently of the background. It might do so in an internal validation dataset but, to our surprise, fails in an external test set - a \textit{problem}. Further investigation reveals that the model acts independently of the animal, and is a good meadow-desert classifier - \textit{an explanation}. This explanation, in some sense, solves the problem: we now understand \textit{why} the model failed on the test dataset. We can use this explanation to design augmentations or generate a new dataset to train a better camel-cow classifier. An explanation should adhere to some standards to be helpful, though.

\textbf{Good explanations } Following the distinction between good and bad explanations by \cite{marcus2024howwhywhen}, an explanation is considered \textit{good} if it adheres to three criteria. Firstly, (1) the explanation should be criticizable. In the case of the camel-cow classifier, we \textit{can} experimentally test the explanation with images of animals in meadows and deserts and view the model's output. Additionally, (2) the explanation should be hard to vary. For the cow-camel classifier, for example, \textit{we can not change} ``meadow'' to ``airplane'' while preserving the explanation's predictive power. Finally, (3) the explanation should be non-authoritarian. In the case of computational pathology, \textit{it should not matter if a famous pathologist, PhD student in machine learning, or a VLM generates the explanation}. For instance, ``The model is a perfect survival predictor because famous pathologist X developed it'' is a bad explanation because it does not explain anything about the reality of how the model functions. In this case, ``the model provides a higher score for more aggressive tumor morphologies'' is a better explanation. 

Given a good explanation, we can start testing its predictive power: Can it accurately predict the output of the black-box model given an image?

\section{An Initial Explanation of Prognosis Prediction} \label{sec:explanation-of-crc}
Now that we have a concept of explanations, we put it into practice. To do so, we test an open-source colorectal cancer risk prediction model by \cite{jiangkathercrcprognostic2024end} for two reasons. Firstly, following posthoc correlative analyses, the authors forge an explanation that reaches similar conclusions to previous work \cite{wulczyn2021interpretable}, which we can challenge. Secondly, survival prediction is a complex task and may provide novel biological insights when presenting a good explanation, as shown by \cite{reitsam2024converging}. 

The model (which we term MIL) is a feature-extraction and multiple instance learning pipeline \cite{ilse2018abmil,schirris2022deepsmile} that 
takes $224 \times 224$ px patches of the tissue from a hematoxylin and eosin stained whole-slide image (WSI) at 1.14 mpp, performs stain normalization \cite{macenko2009method}, extracts features using RetCCL \cite{wang2023retccl}, and classifies an attention-weighted mean of all feature vectors to provide a WSI-level risk logit. See \cite{jiangkathercrcprognostic2024end} for details. Here we will focus on why the model classifies a single patch as high- or low-risk, which is the elementary component of a WSI-level prediction. 

Given the analytical results from \cite{jiangkathercrcprognostic2024end}, we conjecture an explanation: \textit{``MIL assigns higher scores to patches exhibiting: (1) Poorly differentiated or highly proliferative tumor regions, often with epithelial–mesenchymal transition. (2) Invasion into or infiltration of surrounding adipose tissue.
(3) Morphological features indicative of an aggressive tumor-stroma interface. Conversely, lower scores are associated with better differentiation and organized immune infiltration.''}. This explanation aligns with our criteria for good explanations from the last section: It is criticizable, hard-to-vary, and proposed due to evidence from analyses independent of who performed them. 

Now the core problem that we wish to tackle arises: How do we criticize this initial explanation? For example, does MIL indeed predict a lower risk score for the same patch with aggressive tumor morphology if it would have contained immune infiltrate? The transition of morphologies in a WSI provides us a way to perform informal ``interventional'' experiments using sliding windows.

\section{Testing and Falsifying Explanation Components With Intuitive Sliding-Window Experiments} \label{sec:viewer}

\begin{figure}[htb]
\centering
\includegraphics[height=80pt]{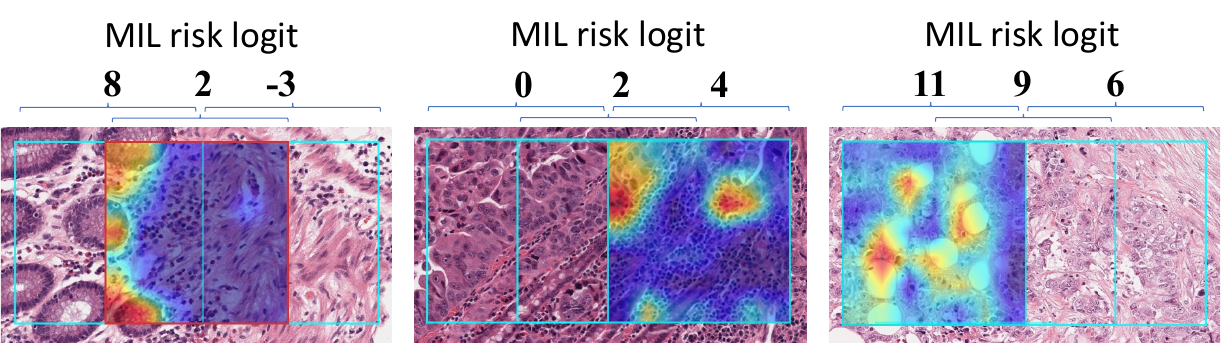}
\caption{Examples of sliding window experiments with an AI-integrated slide viewer. The risk logit as predicted by the MIL model is displayed on top of each patch. The heatmap displays a GradCAM measurement. Left: The right patch receives a low risk logit score, and as glandular features are included to the left, the risk logit increases. Middle: The MIL risk logit increases as lymphocytes are included to the right. Right: As we move to the left and include adipocytes that intermingle with the tumor cells in the right patch, this increases the MIL risk logit further.}
\label{fig:viewer}
\end{figure}

The idea here is to find examples that disagree with, hence falsify, the provided explanation, similar to finding a black swan that falsifies the statement that all swans are white.

We perform sliding window experiments to challenge components of an explanation on overlapping patches from WSI that preserve the original morphology while selectively including or excluding features of interest contained in the explanation. Using our explanation, these experiments assess whether the controlled modifications result in predictable changes in the MIL output. To perform the experiments effectively, we should be able to quickly and intuitively find and select such patches. To this end, we develop a WSI-viewer and integrate the model to perform patch extraction, normalization, inference, and GradCAM analysis on-the-fly. This allows the user to skim over the WSI to find features of interest, especially regions with a gradient of features to capture the change of the MIL predictions as various features are included. 
In Figure \ref{fig:viewer}, we test three components of the explanation: (1) the risk decrease of healthy glands, (2) the risk decrease of lymphocytic infiltration, and (3) the risk increase of tumor cells intermingling with adipocytes. We choose a location in the WSI that contains the histological morphology of the explanation's component we challenge, such that slightly moving a patch would include the other feature, which is done for three overlapping patches. 

\textbf{Healthy glands: } On the left of Fig \ref{fig:viewer}, the black-box risk score increases as normal glandular tissue is included in stromal tissue, confirmed by the GradCAM's highlighting of the glands. This contrasts the explanation in section \ref{sec:definition-explanation}, which states that well-differentiated tissue leads to a lower risk score. This is not incidental: We often find that glandular structures receive very high risk scores. Although a resection of a tumor may rarely have healthy glands which are filtered out with the attention mechanism of MIL, this may pose a problem when a slide has a low tumor-to-tissue ratio.

\textbf{Lymphocytic infiltration: } In the middle of Fig \ref{fig:viewer}, the inclusion of lymphocytic infiltration to an area with relatively organized and heterogeneous solid tumor tissue increases the risk score, confirmed by the GradCAM's focus on both tumor and lymphocytes. This contradicts the explanation, stating that immune response decreases the risk score. During the sliding window experiments the authors noticed that in many other WSI locations regions with lymphocytes do receive a low risk score, but this is not consistently so.

\textbf{Adipocyte infiltration: } On the right of Fig \ref{fig:viewer}, we transition from a region of unorganized, pleomorphic tumor cells towards the left, where adipocytes intermingle with the tumor cells. The explanation states that the score would go up, which is confirmed as we include more and more adipocytes. The GradCAM highlights cells that border the adipocytes, suggesting that the proximity of tumor cells to adipocytes increases the score. These results are highly consistent.


To conclude, the existing explanation is already predictive, but these observations hint at incompleteness of the explanation as MIL's output logits do not always predictively change according to the explanation. The presented tool allows researchers to intuitively perform sliding-window experiments to test components of the explanation. Through refutation of existing components or addition of new insights, improved explanations of the model's functioning are attainable.

If a refined explanation is provided after these experiments, however, a new problem arises: How do we test if this refined explanation is more predictive than the original explanation on a larger dataset? Can we systematically quantify this?


\section{Quantitatively Comparing Competing Explanations} \label{method-validation} \label{sec:evaluation}

Here, the idea is to investigate if we can automatically quantify the predictiveness of an explanation without human involvement to omit the requirement of pathologist annotations. 


We test whether an image description combined with an explanation allows a general-purpose VLM to estimate ($\hat{y}$) the output logit of MIL ($y \in \mathbb{R}$). We use the GPT-4o API to provide a description of the histological features in a patch. Next, we input the image description and the putative explanation, and prompt GPT-4o to predict one of four classes: \[\hat{y} \in \{\texttt{high, medium-high, medium-low, low}\} .\] For all metric calculations, the four-way GPT-4o predictions are binarized (any high versus any low). 

The test set $X$ consists of 132 manually selected patches from 15 randomly samples WSIs from TCGA-CRC available from the GDC repository.




Since the explanations mostly distinguish between features that provide very high and very low scores and does not include an explanation for all possible morphologies present in a WSI, we expect it to be insufficient to distinguish between patches with similar scores close to the decision boundary. To this end, we define subsets \( X_{j} = \{ x_i \in X : |y_i| > j \} \), where samples with small MIL outputs (i.e., near the median score, the decision boundary) are removed.

We compute the bootstrapped AUROC for \(X\) (n=132, $55\%$ high MIL risk logit), \(X_1\) (n=81, $54\%$ high MIL risk logit), \(X_3\) (n=29, $72\%$ high MIL risk logit), i.e., the subsets of the dataset with extreme predictions of MIL. We expect GPT-4o with a poor explanation to fail on all these subsets, while a predictive explanation would achieve better performance on the easier datasets with highly diverging MIL risk logits.

\begin{table}[bt]
\caption{Explanations tested in our experiments.}
\label{tab:explanations}
\centering
\fontsize{8}{10}\selectfont
\begin{tabular}{p{0.95\textwidth}}
\toprule
\multicolumn{1}{c}{\textbf{Explicitly Poor Explanations}} \\
\hline

\textbf{(1) cow\_camel:} The presence of a camel provides a high score, while the presence of a cow provides a low score. Absence of either leads to a medium-high or medium-low score.  \\ 
\textbf{(2) tumor\_lymphocyte\_inverse:} Presence of lymphocytes leads to a higher score. Presence of tumor cells leads to a lower score.  \\
\toprule
\multicolumn{1}{c}{\textbf{Expected Predictive Explanations}} \\
\hline


\textbf{(3) tumor\_lymphocyte:} Presence of lymphocytes leads to a lower score. Presence of tumor cells leads to a higher score. \\
\textbf{(4) detailed\_analyses:} The explanation as described in section \ref{sec:explanation-of-crc}. \\
\bottomrule
\end{tabular}

\end{table}



To see if our method with the VLM and metric calculations is consistent and provides expected outputs, we want to see if an explanation that has nothing to do with pathology (cow camel) achieves a random performance. Additionally, to see if the VLM properly follows the explanation, we provide it with two inversed explanations (tumor lymphocyte (inverse)) to see if the resulting AUROC is inversed. Next, to see if these metrics can distinguish between two competing explanations that are expected to be predictive, we compare a very coarse explanation (tumor lymphocyte) with the explanation derived from detailed analyses (Section \ref{sec:explanation-of-crc}). These explanations are shown in Table \ref{tab:explanations}.



{ 
\setlength{\tabcolsep}{5pt}
\renewcommand{\arraystretch}{1}

\begin{table}[b]
\label{tab:xp-all}
\centering
\caption{Bootstrapped AUROC using varying explanations on subsets of the dataset with increasingly divergent MIL risk logits}
\begin{tabular}{l|ccc }
\toprule
& \multicolumn{3}{c}{Data Subset} \\
Explanation & $X$ & $X_1$ & $X_3$ \\
\hline
cow camel & 0.55$\pm$0.05 & 0.55$\pm$0.06 & 0.60$\pm$0.11  \\
tumor lymphocyte inverse & 0.38$\pm$0.06 & 0.31$\pm$0.07 & 0.11$\pm$0.07  \\
\hline
tumor lymphocyte & 0.62$\pm$0.05 & 0.67$\pm$0.07 & 0.77$\pm$0.11  \\
detailed analysis & 0.62$\pm$0.05 & 0.68$\pm$0.06 & 0.79$\pm$0.09  \\
\bottomrule
\end{tabular}
\end{table}
}

\textbf{Explicitly Poor Explanations: }
Using the cow camel explanation, GPT-4o cannot find any of the explanatory features in the image and essentially defaults to randomly predicting either ``medium-high'' or ``medium-low'', resulting in an AUROC near $0.5$; random performance. When using the inverse of the tumor lymphocyte explanation, predictions follow the explanation, leading to worse-than-random performance as shown by AUROC being markedly lower than $0.5$. On both $X$ and $X_1$ the performance is the inverse of the tumor lymphocyte explanation (\(0.38 = 1 - 0.62\), \(0.31 \approx 1-0.67\)), while on \(X_3\) this explanation is highly (inversely) predictive with an AUROC of $0.11$. Since the AUROC over data subsets follows a similar (but inversed) performance pattern as the normal tumor lymphocyte score, the VLM appears to correctly follow the explanation, independent of how ``sensible'' it is.

\textbf{Assumed Predictive Explanations: }
The two explanations conjectured to be good predictors stand out compared to those designed to be poor predictors, as seen by the considerably higher AUROC, which increases as we test it on only samples with very high and low predictions by the MIL approaching an AUROC of $0.8$. The performance is unexpectedly comparable, though. We hypothesize that the coarse explanation is sufficient for our current set-up with coarse binning into ``high risk'' and ``low risk''. Figure \ref{fig:viewer}, rightmost panel, shows that the MIL risk logit increases from 6 to 11 as tumor cells intermingle with adipocytes, which both explanations would predict to be ``very high'' due to the presence of tumor cells.

To conclude, we provide evidence that the evaluation framework is reasonably able to quantitatively distinguish explanations that are conjectured to be better or worse predictors. The VLM generally follows the explanation as shown by the inverse explanations. Questions about the cause of the similarity between the simple tumor lymphocyte explanation and the detailed explanation remain.   This enables researchers to test the improvement of a refined explanation made with, for example, the sliding window experiments presented in Section \ref{sec:viewer}.


\section{Discussion}

A lack of good explanations for DL models in medical image analysis may lead to disastrous results. Although the field of XAI is growing, explanations are rarely tested, criticized, and iterated upon. We argue this is in part due to the absence of ideas and methods for the falsification and quantification of explanations. This limits the individual researcher to confidently provide good explanations to discover and prevent unwanted actions by the model. 

In this work, we propose that providing a good explanation is necessary and feasible, and take a step towards providing initial conceptual and technical tools for explanation falsification and quantification. We describe what an explanation is and what makes a good explanation. Additionally, we propose that intuitive sliding-window experiments with an AI-integrated image viewer can test components of an explanation. Finally, using a VLM we can quantitatively compare competing explanations, although it may not yet be precise enough to identify the most predictive among a set of multiple predictive explanations.

There are some limitations, however. GPT-4o is not perfect for medical image analysis \cite{cai2024assessing4oblood,zhang2024comparative4ophthal}, hence poor predictiveness of an explanation may be due to a limitation of the VLM. Increasingly performant reasoning models will likely bridge this gap soon, though. Second, we only test one model on a single small dataset for which we already had a good explanation from other analyses. Extending this to a larget dataset is straightforward, though, since we require patches without labels. Additionally, a good exercise is to apply the sliding window approach to a model while being blind to its task to provide an initial explanation from scratch, and continuously refine it with feedback from the quantitative evaluation, to finally validate the explanation on an external dataset.


In the future, we foresee explanations to be generated and evaluated by a multi-agent system, as shown to work for statistical hypothesis testing and WSI classification \cite{ghezloo2025pathfindermultiagentclassification,huang2025automatedmultiagentpopper}. Using the concepts presented in this work, we imagine an agent that selects patches that challenges the existing explanation, and another agent to refute incorrect components, iterating until a stable and predictive explanation is found.

With testable explanations that continue to improve, AI in the medical domain can be better understood, reducing the likelihood of unwanted behavior in the clinic. As these models solve harder tasks using multimodal data, we believe good explanations may, in the future, provide novel biological hypotheses about, e.g., prognostic or predictive features.

%
%
%
\bibliographystyle{splncs04}
\bibliography{mybib}

\end{document}